\documentclass[12pt,twoside]{article}

\usepackage{amsmath}

\DeclareMathAlphabet\mathfrak{U}{euf}{m}{n}
\DeclareFontFamily{U}{euf}{}
\DeclareFontShape{U}{euf}{m}{n}{
  <5> <6> <7> <8> <9> gen * eufm
  <10> <10.95> <12> <14.4> <17.28> <20.74> <24.88> eufm10
  }{}
\DeclareMathAlphabet\mathbb  {U}{msb}{m}{n}
\DeclareFontFamily{U}{msb}{}
\DeclareFontShape{U}{msb}{m}{n}{
  <5> <6> <7> <8> <9> gen * msbm
  <10> <10.95> <12> <14.4> <17.28> <20.74> <24.88> msbm10
  }{}
 
\def\dslash{\partial \kern-0.52em \raisebox{0.2ex}{/} \kern 0.1em}
\def\one{\mathsf{1}}

\newtheorem{prp}{Proposition}
\newtheorem{dfn}[prp]{Definition}
\def\qed{\leavevmode \hbox to.77778em{\hfil\vrule
	 \vbox to.675em{\hrule width.6em\vfil\hrule}\vrule\hfil}}

\makeatletter
\def\section{\@startsection{section}{1}{\z@}{-3.25ex plus -1ex minus -.2ex}%
            {1.5ex plus .2ex}{\normalfont\bfseries}} 

\def\thebibliography#1{\section*{\refname}\list
  {\@biblabel{\theenumiv}}{\settowidth\labelwidth{\@biblabel{#1}}%
    \leftmargin\labelwidth
    \advance\leftmargin\labelsep
    \footnotesize \parsep=0pt \itemsep=0pt
    \usecounter{enumiv}%
    \let\p@enumiv\@empty
    \def\theenumiv{\arabic{enumiv}}}%
    \def\newblock{\hskip .11em plus.33em minus.07em}%
    \sloppy\clubpenalty4000\widowpenalty4000
    \sfcode`\.=1000\relax}

\def\@citex[#1]#2{\if@filesw\immediate\write\@auxout
        {\string\citation{#2}}\fi
\def\@citea{}\@cite{\@for\@citeb:=#2\do
        {\@citea\def\@citea{,}\@ifundefined
        {b@\@citeb}{{\bf ?}\@warning
        {Citation `\@citeb' on page \thepage \space undefined}}
        {\csname b@\@citeb\endcsname}}}{#1}}
\newif\if@cghi
\def\cite{\@cghitrue\@ifnextchar [{\@tempswatrue
        \@citex}{\@tempswafalse\@citex[]}}
\def\citelow{\@cghifalse\@ifnextchar [{\@tempswatrue
        \@citex}{\@tempswafalse\@citex[]}}
\def\@cite#1#2{{\if@cghi\unskip$\null^{#1}$\else [#1]\fi\if@tempswa\typeout
        {warning: optional citation argument ignored: `#2'} \fi}}
 
\def\ps@hep{\addtolength{\headheight}{8pt}
     \addtolength{\topmargin}{-15pt}\addtolength{\headsep}{15pt}
   \def\@oddhead{\hfil\begin{tabular}{c}\texttt{hep-th/9708071}\\
    (revised version)\end{tabular}}
       \let\@evenhead\@oddhead
       \def\@oddfoot{\hfil\thepage\hfil}\let\@evenfoot\@oddfoot}

\makeatother

\begin{document}

\thispagestyle{hep}

\begin{center}

{\large\bfseries Gauge theories with graded differential Lie algebras}
\bigskip

{\renewcommand{\thefootnote}{\fnsymbol{footnote}}
\textsc{Raimar Wulkenhaar}\footnote{present address: 
   Centre de Physique Th\'eorique, CNRS-Luminy, Case 907, 
   F-13288 Marseille (France) due to grant no.\ D/97/20386 
   of the German Academic Exchange Service (DAAD)}\footnote{e-mail: 
   \texttt{raimar@cpt.univ-mrs.fr} or 
   \texttt{raimar.wulkenhaar@itp.uni-leipzig.de}}}
\bigskip

{\small\itshape 
   Institut f\"ur Theoretische Physik, Universit\"at Leipzig \\
   Augustusplatz 10/11, D--04109 Leipzig, Germany} 
\bigskip 

{\small May 5, 1998}

\end{center}

\vskip 2ex

\begin{abstract}

We present a mathematical framework of gauge theories that is based
upon a skew-adjoint Lie algebra and a generalized Dirac operator, both
acting on a Hilbert space.
\vskip 1ex

\end{abstract}

\vskip 2ex

\section{Introduction}

This paper precises the author's previous article \cite{rw2}, in which
we proposed a mathematical calculus towards gauge field theories based
upon graded differential Lie algebras. Given a skew-adjoint Lie
algebra $\mathfrak{g}$, a representation $\pi$ of $\mathfrak{g}$ on a
Hilbert space $h_0$ as well as an [unbounded] operator $D$ and a
grading operator $\Gamma$ on $h_0$, we developed a scheme providing
connection and curvature forms to build physical actions. The general
part of our exposition was on a very formal level, we worked with
unbounded operators (even the splitting of a bounded in two unbounded
operators) without specification of the domain.

In the present paper, we correct this shortcoming. The idea is to
introduce a second Hilbert space $h_1$, which is the domain of the
unbounded operator $D$.  Now, $D$ is a linear continuous operator from
$h_1$ to $h_0$, and the just mentioned splitting involves continuous
operators only. Moreover, the awkward connection theory in the
previous paper is resolved in a strict algebraic description in terms
of normalizers of graded Lie algebras.  Finally, our construction of
the universal graded differential Lie algebra is considerably
simplified (thanks to a hint by K.~Schm\"udgen).

The scope of our framework is the construction of Yang--Mills--Higgs
models in noncommutative geometry \cite{ac}. The standard procedure
\cite{iks,mgv} starts from spectral triples with real structure
\cite{acr,acg} and is limited to the standard model \cite{lmms}. The
hope is \cite{rw1} that the replacement of the unital associative
$*$-algebra in the prior Connes--Lott prescription \cite{cl} by a
skew-adjoint Lie algebra admits representations general enough to
construct grand unified theories. For a realization of this strategy
see refs.\ \citelow{rw3,rw4,rw5}. We discuss the relation to the
axiomatic formulation \cite{acg} of noncommutative geometry in the
last section.
					       
\section{The algebraic setting}

Let $\mathfrak{g}$ be a skew-adjoint Lie algebra, $a^*=-a$ for all $a
\in \mathfrak{g}$. Let $h_0,h_1$ be Hilbert spaces, where $h_1$ is
dense in $h_0$.  Denoting by $\mathcal{B}(h_0)$ and $\mathcal{B}(h_1)$
the algebras of linear continuous operators on $h_0$ and $h_1$,
respectively, we define $\mathcal{B}:=\mathcal{B}(h_0) \cap
\mathcal{B}(h_1)$. The vector space of linear continuous mappings from
$h_1$ to $h_0$ is denoted by $\mathcal{L}$.  Let $\pi$ be a
representation of $\mathfrak{g}$ in $\mathcal{B}$. Let $D \in
\mathcal{L}$ be a generalized Dirac operator with respect to
$\pi(\mathfrak{g})$. This means that $D$ has an extension to a
selfadjoint operator on $h_0$, that $[D,\pi(a)] \in \mathcal{L}$ even
belongs to $\mathcal{B}$ for any $a \in \mathfrak{g}$ and that the
resolvent of $D$ is compact. Finally, let $\Gamma \in \mathcal{B}$ be
a grading operator, i.e.\ $\Gamma^2$ is the identity on both $h_0$ and
$h_1$, $[\Gamma,\pi(a)]=0$ on both $h_0,h_1$ for any $a \in
\mathfrak{g}$ and $D \Gamma + \Gamma D=0$ on $h_1$ extends to $0$ on
$h_0$. This setting was called \emph{L-cycle} in ref.\ \citelow{rw2},
referring to a \emph{Lie}-algebraic version of a \emph{K-cycle}, the
former name \cite{ac,cl} for spectral triple \cite{acr,acg}.

The standard example of this setting $(\mathfrak{g},h_0,h_1,D,\pi,\Gamma)$ 
is
\begin{align}
\mathfrak{g} &= C^\infty(X) \otimes \mathfrak{a}\,, & 
h_0 &= L^2(\mathcal{S}) \otimes \mathbb{C}^F\,, \notag \\ 
h_1 &= W^2_1(\mathcal{S}) \otimes \mathbb{C}^F\,, & 
D &=\mathrm{i} \dslash \otimes \one_F 
+ \boldsymbol{\gamma} \otimes \mathcal{M}\,, \label{standard} \\ 
\pi &=\mathrm{id} \otimes \hat{\pi}\,, & 
\Gamma &= \boldsymbol{\gamma} \otimes \hat{\Gamma}\,. \notag
\end{align}
Here, $C^\infty(X)$ denotes the algebra of real-valued smooth
functions on a compact Riemannian spin manifold $X$, $\mathfrak{a}$ is a
skew-adjoint matrix Lie algebra, $L^2(\mathcal{S})$ denotes the Hilbert
space of square integrable sections of the spinor bundle $\mathcal{S}$
over $X$, $W_1^2(\mathcal{S})$ denotes the Sobolev space of square
integrable sections of $\mathcal{S}$ with generalized first derivative,
$\mathrm{i} \dslash$ is the Dirac operator of the spin connection,
$\boldsymbol{\gamma}$ is the grading operator on $L^2(\mathcal{S})$
anti-commuting with $\mathrm{i} \dslash$, $\hat{\pi}$ is a
representation of $\mathfrak{a}$ in $\mathrm{M}_F \mathbb{C}$ and
$\hat{\Gamma}$ a grading operator on $\mathrm{M}_F \mathbb{C}$ commuting
with $\hat{\pi}(\mathfrak{a})$ and anti-commuting with $\mathcal{M} \in
\mathrm{M}_F \mathbb{C}$.

\section{The universal graded differential Lie algebra $\Omega$}

For $\mathfrak{g}$ being a real Lie algebra we consider the real vector
space $\mathfrak{g}^2=\mathfrak{g} \times \mathfrak{g}$, with the linear
operations given by $\lambda_1 (a_1,a_2)+ \lambda_2 (a_3,a_4) =
(\lambda_1 a_1 + \lambda_2 a_3,\lambda_1 a_2 + \lambda_2 a_4)$, for
$a_i \in \mathfrak{g}$ and $\lambda_i \in \mathbb{R}$. Let $T$ be the
tensor algebra of $\mathfrak{g}^2$, equipped with the $\mathbb{N}$--grading
structure $\deg((a,0))=0$ and $\deg((0,a))=1$, and linear extension to
higher degrees, $\deg(t_1 \otimes t_2)=\deg(t_1) + \deg(t_2)$, for
$t_i \in T$. Defining $T^n=\{t \in T: \deg(t)=n\}$, we have
$T=\bigoplus_{n \in \mathbb{N}} T^n$ and $T^k \otimes T^l \subset
T^{k+l}$. We regard $T$ as a graded Lie algebra with graded commutator
given by $[t^k,t^l]:=t^k \otimes t^l-(-1)^{kl} \, t^l \otimes t^k$,
for $t^n \in T^n$.

Let $\tilde{\Omega}=\bigoplus_{n \in \mathbb{N}} \tilde{\Omega}^n=\sum\,
[\mathfrak{g}^2, [\dots [\mathfrak{g}^2, \mathfrak{g}^2 ] \dots ]]$ be the
$\mathbb{N}$-graded Lie subalgebra of $T$ [due to the graded Jacobi
identity] given by the set of sums of repeated commutators of elements
of $\mathfrak{g}^2$. Let $I'$ be the vector subspace of $\tilde{\Omega}$
of sums of elements of the following type:
\begin{align}
[(a,0),(b,0)] -([a,b],0)~, && [(a,0),(0,b)]+[(0,a),(b,0)] -(0,[a,b])~, 
\end{align}
for $a,b \in \mathfrak{g}$. The first part extends the Lie algebra
structure of $\mathfrak{g}$ to the first component of $\mathfrak{g}^2$
and the second part plays the r\^ole of a Leibniz rule, see
below. Obviously, $I:=I' + [\mathfrak{g}^2,I'] + [\mathfrak{g}^2,
[\mathfrak{g}^2,I']] + \dots $ is an $\mathbb{N}$-graded ideal of
$\tilde{\Omega}$, so that $\Omega :=\bigoplus_{n \in \mathbb{N}}
\Omega^n$, $\Omega^n :=\tilde{\Omega}^n / (I \cap \Omega^n)$ is an
$\mathbb{N}$-graded Lie algebra.

On $T$ we define recursively a graded differential as an
$\mathbb{R}$-linear map $d: T^n \to T^{n+1}$ by
\begin{align}
d (a,0) &= (0,a)~, & d (0,a) &= 0~, \label{datt} \\ d ( (a,0) \otimes
t) &= d(a,0) \otimes t + (a,0) \otimes d t ~, & d ( (0,a) \otimes t)
&= - (0,a) \otimes d t ~, \notag
\end{align}
for $a \in \mathfrak{g}$ and $t \in T$. One easily verifies $d^2=0$ on
$T$ and the graded Leibniz rule $d(t^k \otimes t^l)=dt^k \otimes t^l +
(-1)^k t^k \otimes dt^l$, for $t^n \in T^n$. Thus, $d$ defined by
\eqref{datt} is a graded differential of the tensor algebra $T$ and of
the graded Lie algebra $T$ as well, $d[t^k,t^l] = [dt^k,t^l] + (-1)^k
[t^k,dt^l]$.

Due to $d \mathfrak{g}^2 \subset \mathfrak{g}^2$ we conclude that $d$ is
also a graded differential of the graded Lie subalgebra
$\tilde{\Omega} \subset T$.  Moreover, one easily checks $dI' \subset
I'$, giving $dI \subset I$.  Therefore, $(\Omega\,,\,[~,~]\,,\,d)$ is
a graded differential Lie algebra, with the graded differential $d$
given by $d( \varpi+I):= d \varpi + I$, for $\varpi \in
\tilde{\Omega}$.

We extend the involution $*: a \mapsto -a$ on $\mathfrak{g}$ to an
involution of $T$ by $(a,0)^*=-(a,0)$, $(0,a)^*=-(0,a)$ and $(t_1
\otimes t_2)^*= t_2^* \otimes t_1^*$, giving $[t^k,t^l]^*=-(-1)^{kl}
[t^k{}^*,t^l{}^*]$. Clearly, this involution extends to $\Omega$. The
identity $a=-a^*$ yields $\omega^k{}^* = -(-1)^{k(k-1)/2} \omega^k$,
for any $\omega^k \in \Omega^k$.

The graded differential Lie algebra $\Omega$ is universal in the
following sense:
\begin{prp} 
\label{upr}
Let $\Lambda=\bigoplus_{n \in \mathbb{N}} \Lambda^n$ be an
$\mathbb{N}$-graded Lie algebra with graded differential $\mathrm{d}$
such that \\ \hbox to 1.5em {\hss \textup{i)}}
$~\Lambda^0=\pi(\mathfrak{g})$ for a surjective homomorphism $\pi$ of
Lie algebras, \\ \hbox to 1.5em {\hss \textup{ii)}} $~\Lambda$ is
generated by $\pi(\mathfrak{g})$ and $\mathrm{d}\pi(\mathfrak{g})$ as
the set of repeated commutators.  \\ Then there exists a differential
ideal $I_\Lambda \subset \Omega$ fulfilling $\Lambda \cong
\Omega/I_\Lambda$.
\end{prp}
\noindent
\textit{Proof.} Define a linear surjective mapping $p:\Omega \to
\Lambda$ by
\begin{align*}
p((a,0)) &= \pi(a)~, &
p(d \omega) &= \mathrm{d}(p(\omega))~, &
p([\omega,\tilde{\omega}]) &:= [p(\omega),p(\tilde{\omega})] ~,
\end{align*}
for $a \in \mathfrak{g}$ and $\omega,\tilde{\omega} \in \Omega$.  Because
of $d \ker p \subset \ker p$, $I_\Lambda =\ker p$ is the desired
differential ideal of $\Omega$. \hfill \qed

\section{The graded differential Lie algebra $\Omega_D$} 
			   
Using the grading operator $\Gamma$ we define on $\mathcal{L}$ and
$\mathcal{B}$ a $\mathbb{Z}_2$--grading structure, the even subspaces
carry the subscript $0$ and the odd subspaces the subscript $1$. Then,
the graded commutator $[\,.\,,\,.\,]_g : \mathcal{L}_i \times
\mathcal{B}_j \to \mathcal{L}_{(i{+}j)\:\mathrm{mod}\: 2}$ is defined
by
\begin{equation}
[ A_i ,B_j]_g := A_i \circ B_j -(-1)^{ij} B_j \circ A_i 
\equiv -(-1)^{ij} [ B_j ,A_i]_g ~,
\end{equation}
where $B_j \in \mathcal{B}_j$ and $A_i \in \mathcal{L}_i$. If $A_i \in
\mathcal{B}_i$ then $[\,.\,,\,.\,]_g$ maps $h_1$ to $h_1$ and $h_0$
to $h_0$.

Using the elements $\pi$ and $D$ of our setting we define a linear
mapping $\pi: \Omega \to \mathcal{B}$ by
\begin{equation}
\begin{split}
\pi( (a,0)) &:= \pi(a)~, \qquad \qquad \qquad
\pi( (0,a)) := [-\mathrm{i} D,\pi(a)]_g ~,\\
\pi([\omega^k,\omega^l]) &:= [\pi(\omega^k),\pi(\omega^l)]_g ~, 
\end{split} \label{pi}
\end{equation}
for $a \in \mathfrak{g}$ and $\omega^n \in \Omega^n$. The selfadjointness
of $D$ on $h_0$ implies that $\pi$ is involutive,
$\pi(\omega^*)=(\pi(\omega))^*$.

Note that $\pi(\Omega)$ is not a graded \emph{differential} Lie
algebra. The standard procedure to construct such an object is to
define $\mathcal{J} = \ker\pi + d \ker \pi \subset \Omega$. It is easy to
show that $\mathcal{J}$ is a graded differential ideal of $\Omega$,
providing the graded differential Lie algebra
\begin{align}
\Omega_D &= {\textstyle \bigoplus_{n \in \mathbb{N}}} \; \Omega_D^n~,& 
\Omega_D^n := \Omega^n / \mathcal{J}^n 
\cong \pi(\Omega^n) / \pi(\mathcal{J}^n) ~,
\end{align}
where $\mathcal{J}^n = \mathcal{J} \cap \Omega^n$. One has $\Omega_D^0
\cong \pi(\Omega^0) \equiv \pi(\mathfrak{g})$ and $\Omega_D^1 \cong
\pi(\Omega^1)$. By construc\-tion, the differential $d$ on $\Omega_D$
is given by $d (\pi(\omega^n) + \pi(\mathcal{J}^n )) := \pi(d\omega^n)
+\pi(\mathcal{J}^{n+1})$, for $\omega^n \in \Omega^n$.
	 
It is very useful to consider an extension of the second formula of
\eqref{pi}, $\pi( d(a,0)) := [-\mathrm{i} D,\pi((a,0))]_g$, to higher
degrees:
\begin{equation}
\pi( d \omega^n ) = [-\mathrm{i} D,\pi(\omega^n)]_g +
\sigma(\omega^n)~, \quad \omega^n \in \Omega^n~. \label{pdo}
\end{equation}
It turns out \cite{rw2} that $\sigma: \Omega \to \mathcal{L}$ is a linear
mapping recursively given by
\begin{equation}
\begin{split}
\sigma( (a,0)) &= 0 ~,\qquad \qquad \qquad
\sigma( (0,a)) = [D,[D,\pi(a)]_g]_g ~, \\
\sigma([\omega^k,\omega^l]) &= [\sigma(\omega^k),\pi(\omega^l)]_g 
+ (-1)^k [\pi(\omega^k),\sigma(\omega^l)]_g ~.
\end{split} \label{sigma}
\end{equation}
Equation \eqref{pdo} has an important consequence: Putting $\omega^n
\in \ker \pi$ we get
\begin{equation}
\pi(\mathcal{J}^{n+1})= \{ \sigma(\omega^n)~:~~ \omega^n \in \Omega^n
\cap \ker \pi \}~.
\end{equation}
The point is that $\sigma(\Omega)$ can be computed from the last
equation \eqref{sigma} once $\sigma(\Omega^1)$ is known. Then one can
compute $\pi(\mathcal{J})$ and obtains with \eqref{pdo} an explicit
formula for the differential on $\Omega_D$.

\section{Connections}

We define the graded normalizer $N_{\mathcal{L}}(\pi(\Omega))$ of
$\pi(\Omega)$ in $\mathcal{L}$, its vector subspace $\mathcal{H}$
compatible with $\pi(\mathcal{J})$ and the graded centralizer
$\mathcal{C}$ of $\pi(\Omega)$ in $\mathcal{L}$ by
\begin{align}
N^k_{\mathcal{L}}(\pi(\Omega)) &= 
\{ \eta^k \in \mathcal{L}_{k\,\mathrm{mod}\,2} \,:\: 
\text{$\eta^k$ has $\!\genfrac{\{}{\}}{0pt}{1}{\text{selfadjoint}}{
\text{skew-adjoint}}\!$ extension for $\tfrac{k(k{-}1)}{2}$
$\!\genfrac{\{}{\}}{0pt}{1}{\text{odd}}{\text{even}}$}\,, \notag \\ &
{} \quad\quad [\eta^k , \pi(\omega^n) ]_g \in \pi(\Omega^{k+n})~~ 
\forall \omega^n \in \Omega^n\,,~ \forall n \in \mathbb{N} \, \}\,,
\notag \\ 
\mathcal{H}^k &= \{ \eta^k \in N^k_{\mathcal{L}}(\pi(\Omega))\;:~
[\eta^k, \pi(j^n)]_g \in \pi(\mathcal{J}^{k+n})~~ 
\forall j^n \in \mathcal{J}^n \, \}\,, \\ 
\mathcal{C}^k &= \{ c^k \in N^k_{\mathcal{L}}(\pi(\Omega))\;:~ 
[c^k , \pi(\omega) ]_g =0 ~~\forall \omega \in \Omega \, \}\,. 
\notag
\end{align}
Here, the linear continuous operator $[\eta^k , \pi(\omega^n) ]_g :h_1
\to h_0$ must have its image even in the subspace $h_1 \subset h_0$
and must have an extension to a linear continuous operator on $h_0$.
For each degree $n$ we have the following system of inclusions:
\begin{equation}
\begin{array}{ccccccc}
\mathcal{L} & \supset & \mathcal{H}^n & \supset & \!\pi(\Omega^n)\! &
\supset & \!\pi(\mathcal{J}^n)\! \\ 
&& \!\cup\, && \cap && \\ 
&& \mathcal{C}^n && \mathcal{B} & \subset & \mathcal{L}
\end{array}
\end{equation}
The graded Jacobi identity and Leibniz rule define the structure of a
graded differential Lie algebra on $\hat{\mathcal{H}} = \bigoplus_{n
\in \mathbb{N}}\, \hat{\mathcal{H}}^n,$ with $\hat{\mathcal{H}}^n
=\mathcal{H}^n/(\mathcal{C}^n {+} \pi(\mathcal{J}^n))$:
\begin{align}
[[\eta^k {+} \mathcal{C}^k {+} \pi(\mathcal{J}^k) & \,,\, 
\eta^l {+} \mathcal{C}^l {+} \pi(\mathcal{J}^l)]_g \,,\, 
\pi(\omega^n) {+} \pi(\mathcal{J}^n)]_g
\notag \\ & 
\hspace*{-3em} := [\eta^k, [\eta^l , \pi(\omega^n)]_g ]_g
- (-1)^{kl} [\eta^l, [\eta^k , \pi(\omega^n)]_g ]_g 
+ \pi(\mathcal{J}^{k+l+n}) \,, \notag \\{} 
[ d (\eta^k {+} \mathcal{C}^k {+} \pi(\mathcal{J}^k)) & 
\,,\, \pi(\omega^n) {+} \pi(\mathcal{J}^n)]_g \label{jaclei} \\
& \hspace*{-3em} := \pi \circ d \circ \pi^{-1} ( [ \eta^k,
\pi(\omega^n)]_g) - (-1)^k [ \eta^k, \pi(d \omega^n)]_g 
+ \pi(\mathcal{J}^{k+n+1}) \,, \notag
\end{align}
for $\eta^n \in \mathcal{H}^n$ and $\omega^n \in \Omega^n$. 

The lesson is that $\pi(\Omega)$ and its ideal $\pi(\mathcal{J})$ give
rise not only to the graded differential Lie algebra $\Omega_D$ but
also to $\hat{\mathcal{H}}$, both being natural. It turns out that it
is the differential Lie algebra $\hat{\mathcal{H}}$ which occurs in
our connection theory:
\begin{dfn}
\label{def1}
Within our setting, a connection $\nabla$ and its associated covariant
derivative $\mathcal{D}$ are defined by \\[0.5ex]
\hbox to 2em{\hss \textup{i)}} $~\mathcal{D} \in \mathcal{L}_1$ with
selfadjoint extension, \\
\hbox to 2em{\hss \textup{ii)}} $~\nabla: \Omega_D^n \to \Omega_D^{n+1}$ 
is linear, \\
\hbox to 2em{\hss \textup{iii)}} $~\nabla (\pi(\omega^n) {+} 
\pi(\mathcal{J}^n)) = [-\mathrm{i} \mathcal{D}, \pi(\omega^n)]_g 
+ \sigma(\omega^n) + \pi(\mathcal{J}^{n+1})~, \quad \omega^n \in \Omega^n$.
\\[0.5ex]
The operator $\nabla^2: \Omega_D^n \to \Omega_D^{n+2}$ is called the
curvature of the connection. 
\end{dfn} 
This definition states that the covariant derivative $\mathcal{D}$
generalizes the operator $D$ of the setting and the connection
$\nabla$ generalizes the differential $d$. In particular, both
$\mathcal{D}$ and $\nabla$ are related via the same equation iii) as
$D$ and $d$ are according to \eqref{pdo}.
\begin{prp}
Any connection/covariant derivative has the form
$\nabla=d+[\rho{+}\mathcal{C}^1\,,~.~]_g$ and $\mathcal{D}= D +
\mathrm{i} \rho$, for $\rho \in \mathcal{H}^1$. The curvature is
$\nabla^2=[\theta,~.~]$, with $\theta=d \hat{\rho} + \tfrac{1}{2}
[\hat{\rho},\hat{\rho}]_g \in \hat{\mathcal{H}}^2$, where
$\hat{\rho}=\rho+\mathcal{C}^1 \in \hat{\mathcal{H}}^1$.
\end{prp}
\textit{Proof.} There is a canonical pair of connection/covariant
derivative given by $\nabla=d$ and $\mathcal{D}=D$. If
$(\nabla^{(1)},\mathcal{D}^{(1)})$ and
$(\nabla^{(2)},\mathcal{D}^{(2)})$ are two pairs of
connections/covariant derivatives, we get from iii) \\[0.5ex]
\centerline{$(\nabla^{(1)}{-}\nabla^{(2)})(\pi(\omega^n) {+}
\pi(\mathcal{J}^n))=[\nabla^{(1)}_h {-} \nabla^{(2)}_h,
\pi(\omega^n)]_g + \pi(\mathcal{J}^{n+1})$.}  \\[0.5ex] This means
that $\rho := \nabla^{(1)}_h {-} \nabla^{(2)}_h \in \mathcal{H}^1$ is
a concrete representative and $\nabla^{(1)}{-}\nabla^{(2)} =
[\hat{\rho},\,.\,]_g$, where $\hat{\rho}=\rho+\mathcal{C}^1 \in
\hat{\mathcal{H}}^1$. The formula for $\theta$ is a direct consequence
of \eqref{jaclei}. \hfill \qed

\section{Gauge transformations} 
 
The exponential mapping defines a unitary group 
\begin{align}
\textstyle 
\mathcal{U}:= \{ \prod_{\alpha=1}^N \exp(v_{\alpha}) :~~ & \textstyle
\exp(v_\alpha) := \one_{\mathcal{B}} + \sum_{k=1}^\infty \frac{1}{k!} 
(v_{\alpha})^k\,, \notag \\
& v_\alpha \in \mathcal{H}^0 \cap \mathcal{B}\,,~ 
d v_\alpha -[-\mathrm{i} D,v_\alpha] \in \mathcal{C}^1\; \}\,. \label{gg}
\end{align}
Due to $\exp(v) A \exp(-v) = A+ \sum_{k=1}^\infty \frac{1}{k !}
([v,[v, \dots ,[v , A] \dots ]])_k$, where $(~~)_k$ contains $k$
commutators of $A \in \mathcal{L}$ with $v$, we have a natural
degree-preserving representation $\mathrm{Ad}$ of $\mathcal{U}$ on
$\mathcal{H}$, $\mathrm{Ad}_u (\eta^n)= u \eta^n u^* \in
\mathcal{H}^n$, for $\eta^n \in \mathcal{H}^n$ and $u \in
\mathcal{U}$.
\begin{dfn} 
In our setting, the gauge group is the group $\mathcal{U}$ defined in 
\eqref{gg}. Gauge transformations of the connection and the covariant
derivative are given by
\\[0.5ex]
\centerline{ 
$\nabla \mapsto \nabla' := \mathrm{Ad}_u \nabla \mathrm{Ad}_{u^*}~,
\quad \mathcal{D} \mapsto \mathcal{D}' := u \mathcal{D}u^*~, \quad 
u \in \mathcal{U}$. } 
\end{dfn} 
Note that the consistency relation iii) in Definition~\ref{def1}
reduces on the infinitesimal level to the condition $d v_\alpha
-[-\mathrm{i} D,v_\alpha] \in \mathcal{C}^1$ in \eqref{gg}. The gauge
transformation of the curvature form  reads $\theta \mapsto
\theta'=\mathrm{Ad}_{u} \theta$.

\section{Physical action}
\label{pa}

We borrow the integration calculus introduced by Connes to
noncommutative geometry \cite{ac,acr} and summarize the main
results. Let $E_n$ be the eigenvalues of the compact operator
[compactness was assumed in the setting] $|D|^{-1}=(D D^*)^{-1/2}$ on
$h_0$, arranged in decreasing order.  Here, the finite dimensional
kernel of $D$ is not relevant so that $E_1 < \infty$. The K-cycle
$(h_0,D)$ over the $C^*$-algebra $\mathcal{B}(h_0)$ is called
$\mathrm{d}^+$-summable if $\sum_{n=1}^N E_n = O(\, \sum_{n=1}^N
n^{-1/\mathrm{d}} \,)$. Equivalently, the partial sum of the first $N$
eigenvalues of $|D|^{-\mathrm{d}}$ has [at most] a logarithmic
divergence as $N \to \infty$ so that $|D|^{-\mathrm{d}}$ belongs to
the [two-sided] Dixmier trace class ideal
$\mathcal{L}^{(1,\infty)}(h_0)$. Therefore, $f\,|D|^{-\mathrm{d}} \in
\mathcal{L}^{(1,\infty)}(h_0)$ for any $f \in \mathcal{B}(h_0)$, and
the Dixmier trace provides for $f >0$ a linear functional $f \mapsto
\mathrm{Tr}_\omega (f\,|D|^{-\mathrm{d}}) = \mathrm{Lim}_\omega \,
\frac{1}{\ln N} \sum_{n=1}^N \mu_n \in \mathbb{R}^+$. Here, $\mu_n$
are the eigenvalues of $f\,|D|^{-\mathrm{d}}$ and the limit
$\mathrm{Lim}_\omega$ involves an appropriate limiting procedure
$\omega$. The Dixmier trace fulfills $\mathrm{Tr}_\omega
(f\,|D|^{-\mathrm{d}}) =\mathrm{Tr}_\omega
(ufu^*\,|D|^{-\mathrm{d}})$, for unitary $u \in \mathcal{B}(h_0)$.

Let $\theta^*_0 : h_0 \to h_1$ be the uniquely determined adjoint of a
representative $\theta_0: h_1 \to h_0$ of the curvature form $\theta
\in \hat{\mathcal{H}}^2$. It follows $\theta_0 \theta_0^* \in
\mathcal{B}(h_0)$ so that we propose the following definition for the
physical action:
\begin{dfn}
The bosonic action $S_B$ and the fermionic action $S_F$ of the
connection $\nabla$ and covariant derivative $\mathcal{D}$ are given by
\begin{equation}
\begin{split}
S_B(\nabla) &:= \min_{\mathrm{j}^2 \in \mathcal{C}^2+\pi(\mathcal{J}^2)}
\mathrm{Tr}_\omega ((\theta_0 {+} \mathrm{j}^2)(\theta_0 {+}
\mathrm{j}^2)^* |D|^{-\mathrm{d}}) ~, 
\\ 
S_F(\psi,\mathcal{D}) &:=\langle \psi ,\mathcal{D} \psi \rangle_{h_0} ~, 
\quad \psi \in h_1~,
\end{split}
\label{act}
\end{equation}
where $\langle~,~\rangle_{h_0}$ is the scalar product on $h_0$.
\end{dfn}
The bosonic action $S_B$ is independent of the choice of the
representative $\theta_0$. Thus, we can take the canonical
dependence of the gauge potential $\rho$, 
\[
\theta_0 = \{- \mathrm{i} D,\rho\} + \tfrac{1}{2} \{\rho,\rho\} 
+ \sigma \circ \pi^{-1}(\rho)~,  
\]
where $\sigma \circ \pi^{-1}$ is supposed to be extended from
$\pi(\Omega^1)$ to $\mathcal{H}^1$. It is unique up to elements of
$\mathcal{C}^2 {+} \pi(\mathcal{J}^2)$. Since the Dixmier trace is
positive, the element $\mathrm{j}_0^2 \in \mathcal{C}^2
{+}\pi(\mathcal{J}^2)$ at which the minimum in \eqref{act} is attained
is the unique solution of the equation
\[
\mathrm{Tr}_\omega ((\theta_0 {+} \mathrm{j}^2_0) 
(\mathrm{j}^2)^* |D|^{-\mathrm{d}}) =0 ~,  \quad \forall\, \mathrm{j}^2
\in \mathcal{C}^2 {+} \pi(\mathcal{J}^2)~. 
\]

It is clear that the action \eqref{act} is invariant under gauge 
transformations
\begin{equation}
\nabla \mapsto \mathrm{Ad}_u \nabla \mathrm{Ad}_{u^*} ~,\quad 
\mathcal{D} \to u \mathcal{D} u^* ~, \quad \psi \mapsto u \psi~, 
\quad u \in \mathcal{U}~.
\end{equation}
Note that our gauge group as defined in \eqref{gg} is always
connected, which means that we have no access to `big' gauge
transformations.  Note further that there exist Lie groups having the
same Lie algebra. In that case there will exist fermion multiplets
$\psi$ which can be regarded as multiplets of different Lie groups.
For the bosonic sector only the Lie algebra is important, so one can
have the pathological situation of a model with identical particle
contents and identical interactions, but different gauge groups. We
consider such gauge theories as identical.

\section{Remarks on the standard example}

Recall \eqref{pi} that the general form of an element 
$\tau^1 \in \pi(\Omega^1)$ is 
\begin{equation}
\tau^1=\sum_{\alpha,z\geq 0} [\pi(a^z_\alpha),[\dots [\pi(a^1_\alpha),
[-\mathrm{i} D,\pi(a^0_\alpha)]] \dots ]]~, \qquad a^i_\alpha \in 
\mathfrak{g}~.
\end{equation}
For $a^i_\alpha=f^i_\alpha \otimes \hat{a}^i_\alpha \in C^\infty(X)
\otimes \mathfrak{a}$ we get with \eqref{standard}
\begin{equation}
\begin{split}
\tau^1 = \sum_{\alpha,z\geq 0} \Big( & f^z_\alpha \cdots f^1_\alpha
\dslash(f^0_\alpha) \otimes \hat{\pi}([\hat{a}^z_\alpha, [\dots
[\hat{a}^1_\alpha,\hat{a}^0_\alpha] \dots ]]) \\ + & f^z_\alpha \cdots
f^1_\alpha f^0_\alpha \boldsymbol{\gamma} \otimes
[\hat{\pi}(\hat{a}^z_\alpha),[\dots [\hat{\pi}(\hat{a}^1_\alpha),
[-\mathrm{i} \mathcal{M},\hat{\pi}(\hat{a}^0_\alpha)]] \dots ]] \Big)~.
\end{split}
\label{tau1}
\end{equation}
Let us first assume that $\mathfrak{a}$ is semisimple. In this case
the two lines in \eqref{tau1} are independent.  The first line belongs
to $\Lambda^1 \otimes \hat{\pi}(\mathfrak{a})$, because the gamma
matrices occurring in $\dslash$ provide a 1-form basis. In physical
terminology, these Lie algebra-valued 1-forms are Yang--Mills fields
acting via the representation $\mathrm{id}\otimes \hat{\pi}$ on the
fermions. In the second line of \eqref{tau1} we split $\mathcal{M}$
into generators of irreducible representations of $\mathfrak{a}$,
tensorized by generation matrices. Obviously, these irreducible
representations are spanned after taking the commutators with
$\hat{\pi}(\hat{a}^i_\alpha)$. Thus, the second line of \eqref{tau1}
contains sums of function-valued representations of the matrix Lie
algebra [times $\boldsymbol{\gamma}$ and generation matrices], which
are physically interpreted as Higgs fields. In other words, the
prototype $\tau^1$ of a connection form (=gauge potential) describes
representations of both Yang--Mills and Higgs fields on the fermionic
Hilbert space.

{}From a physical point of view, this is a more satisfactory picture
than the usual noncommutative geometrical construction of
Yang--Mills--Higgs models \cite{iks,mgv}. Namely, descending from
Connes' noncommutative geometry \cite{ac,acr,acg} there is only a
limited set of Higgs multiplets possible \cite{tk}: Admissible Higgs
multiplets are tensor products $\mathbf{n} \otimes \mathbf{m}^*$ of
fundamental representations (and their complex conjugate)
$\mathbf{n},\mathbf{m}$ of simple gauge groups, where the adjoint
representation never occurs. This rules out \cite{lmms} the
construction of interesting physical models. In our framework there
are no such restrictions and~-- depending on the choice of
$\mathcal{M}$ and $h$~-- Higgs fields in any representation of a Lie
group are possible. Thus, a much larger class of physical models can
be constructed.

The treatment of Abelian factors $\mathfrak{a}'' \subset \mathfrak{a}$
in our approach is somewhat tricky. One remarks that in the first line
of \eqref{tau1} only the $(z{=}0)$-component of $\mathfrak{a}''$
survives. The consequence is that linear independence of the two lines
in \eqref{tau1} is not automatical. Thus, to avoid pathologies, we
need a condition \cite{rw2} between $\mathcal{M}$ and the
representations of $\mathfrak{a}$ to assure independence. The
$\mathrm{u(1)}$-part of the standard model is admissible in this
sense.

The second consequence of the missing $(z{>}0)$-components in the
first line is that the spacetime $1$-form part of Abelian factors in
$\tau^1$ is a total differential $\dslash(f^0_0) \in d \Lambda^0
\subset \Lambda^1$. This seems to be a disaster at first sight for the
description of Abelian Yang--Mills fields. However, our gauge
potential lives in the bigger space $\mathcal{H}^1 \supset
\pi(\Omega^1)$. Always if there is a part $d \Lambda^0 \otimes
\pi(\mathfrak{a}'')$ in $\pi(\Omega^1)$ there is a part $\Lambda^1
\otimes \pi(\mathfrak{a}'')$ in $\mathcal{H}^1$. There can be even
further contributions from $\mathcal{H}^1$ to the gauge potential,
which are difficult to control in general. Fortunately, it turns out
\cite{rw2} that after imposing a locality condition for the connection
(which is equivalent to saying that $\rho$ commutes with functions),
possible additional $\mathcal{H}^1$-degrees of freedom are either of
Yang--Mills type or Higgs type.

This framework of gauge field theories was successfully applied to
formulate the standard model \cite{rw3}, the flipped $\mathrm{SU(5)}
\times \mathrm{U(1)}$-grand unification \cite{rw4} and
$\mathrm{SO(10)}$-grand unification \cite{rw5}. It is not possible to
describe pure electrodynamics. The reason is that in the Abelian case
the curvature form $\theta \neq 0$ commutes with all elements of
$\pi(\Omega)$. Hence, it belongs to the graded centralizer
$\mathcal{C}^2$ and is projected away in the bosonic action
\eqref{act}.

\section{Do the axioms of noncommutative geometry extend to the Lie
algebraic setting?}

The present status of noncommutative geometry is that this theory is
governed by seven axioms \cite{acg}. In the commutative case, these
axioms provide the algebraic description of classical spin
manifolds. The question now is whether or not our Lie algebraic
version, which is in close analogy with the prior Connes--Lott
formulation \cite{cl} of noncommutative geometry, can also be brought
into contact with Connes' axioms. We list and discuss below the axioms
in their form they would have in terms of Lie algebras.

\begin{list}{}{\settowidth{\labelwidth}{9)}
               \leftmargin\labelwidth\addtolength{\leftmargin}{\labelsep}
               \parsep0pt\parskip0pt}

\item[1)] \emph{Dimension: $|D|^{-1}$ is an infinitesimal of order
$\frac{1}{\mathrm{d}}$, i.e.\ the eigenvalues $E_n$ of $|D|^{-1}$ grow
as $n^{-1/\mathrm{d}}$, where $\mathrm{d}$ is an even natural number.}

\item[3)] \emph{Smoothness: For any $a \in \mathfrak{g}$, both $a$ and
$[D,a]$ belong to the domain of $\delta^m$, where $\delta(\;.\;)
:=[|D|,\;.\;]$.}  
\\ 
The axioms 1) and 3) can be directly transferred to the Lie algebraic
setting. We cannot treat the odd-dimensional case as the
grading operator $\Gamma$ is essential to detect the sign for the
graded commutator.

\item[4)] \emph{Orientability:} Connes requires the
$\mathbb{Z}_2$-grading operator $\Gamma$ to be the image under $\pi$
of a Hochschild $\mathrm{d}$-cycle. We are not going to touch the
extension of Hochschild homology to Lie algebras, but even a
requirement such as $\Gamma \in \pi(\Omega^\mathrm{d})$ is
problematic. For the standard example we have the decomposition
$\Gamma=\boldsymbol{\gamma} \otimes \hat{\Gamma}$, and the comparison
with the general form \cite{rw2} of $\pi(\Omega^\mathrm{d})$ yields
that $\hat{\Gamma}$ has to be the image under $\hat{\pi}$ of the
non-abelian part of $\mathfrak{a}$. In all models we have studied so
far this is not the case. It seems to be impossible to maintain
orientability in our framework. The grading operator $\Gamma$, which
commutes with $\pi(\mathfrak{g})$ and anti-commutes with $D$, is an
extra piece which has no relation with orientability.

\item[7)] \emph{Reality: There exists an anti-linear isometry $J: h_i
\to h_i$ such that $[\pi(a),J\pi(b)J^{-1}]=0$ for all $a,b \in
\mathfrak{g}$, $J^2=\epsilon$, $JD=DJ$ and $J \Gamma = \epsilon'
\Gamma J$, with $\epsilon = (-1)^{\mathrm{d}(\mathrm{d}+2)/8}$ and
$\epsilon' = (-1)^{\mathrm{d}/2}$.}

\item[2)] \emph{First order: $[[D,\pi(a)],J\pi(b)J^{-1}]=0$ for all
$a,b \in \mathfrak{g}$.}

Both axioms 7) and 2) can be trivially fulfilled as soon as an
anti-linear involution $\mathcal{I}$ on $h_i$ is available. It
suffices to define
\begin{align*}
J &:= \left( \begin{array}{cc} 0 & \epsilon\, \mathcal{I}^{-1} \\ 
\mathcal{I} & 0 \end{array} \right)~, &
h_i &\mapsto \left( \begin{array}{c} h_i \\ h_i \end{array}\right)~, &
D &\mapsto \left( \begin{array}{cc} D & 0 \\ 
0 & \mathcal{I} D \mathcal{I}^{-1} \end{array} \right)~, \\ &&
\pi(a) & \mapsto \left( \begin{array}{cc} \pi(a) & 0 \\ 0 & 0
\end{array} \right)~, &
\Gamma &\mapsto \left( \begin{array}{cc} \Gamma & 0 \\ 
0 & \epsilon' \mathcal{I} \Gamma \mathcal{I}^{-1}
\end{array} \right)~.
\end{align*}
The question is whether there are nontrivial real structures which
also satisfy the other axioms. The existence of the real structure $J$
(Tomita's involution) is a central piece of Connes' theory. It has
proved very useful in understanding the commuting electroweak and
strong sectors of the standard model. The same idea could be applied
to our formulation of the standard model \cite{rw3}.  For other gauge
theories \cite{rw4,rw5}, however, a nontrivial real structure $J$
seems to be rather disturbing as it requires the fermions to sit in
(generalized) adjoint representations. To achieve this one had to add
auxiliary $\mathrm{u(1)}$-factors, which is in contradiction to the
grand unification philosophy.

\item[5)] \emph{Finiteness and absolute continuity:} Connes requires 
$h_\infty=\bigcap_m \mathrm{domain}(D^m)$ to be a finite
projective module. Thus, our task would be to define the notion of a
finite projective module over a Lie algebra $\mathfrak{g}$ and the
Lie analogues of the $K$-groups. We are not aware of these structures,
but without them it is impossible to talk about generalizations of the
index pairing of $D$ with the $K$-groups and of

\item[6)] \emph{Poincar\'e duality}.
\end{list}

\noindent
In conclusion, our Lie algebraic version of noncommutative geometry is
not a possible generalization of classical spin manifolds, or at least
there is a lot to do to derive the Lie analogues of standard algebraic
structures. Our approach provides a powerful tool to build gauge field
theories with spontaneous symmetry breaking, the price for this
achievement is the lost of any contact with spin manifolds.


\begin{thebibliography}{99}

\bibitem{rw2} R.~Wulkenhaar, \emph{Non-commutative geometry with
graded differential Lie algebras}, J.\ Math.\ Phys.\ \textbf{38}
(1997) 3358--3390.

\bibitem{ac} A.~Connes, \emph{Non commutative geometry}, Academic
Press, New York 1994.

\bibitem{iks} B.~Iochum, D.~Kastler and T.~Sch\"ucker, \emph{Fuzzy
mass relations in the standard model}, preprint
\texttt{hep-th/9507150}.

\bibitem{mgv} C.~P.~Mart\'{\i}n, J.~M.~Gracia--Bond\'{\i}a and
J.~C.~Varilly, \emph{The standard model as a noncommutative geometry:
the low energy regime}, Phys.\ Rep.\ \textbf{294} (1998) 363--406.

\bibitem{acr} A.~Connes, \emph{Noncommutative geometry and reality},
J.\ Math.\ Phys.\ \textbf{36} (1995) 6194--6231.

\bibitem{acg} A.~Connes, \emph{Gravity coupled with matter and the
foundation of noncommutative geometry}, Commun.\ Math.\ Phys.\
\textbf{182} (1996) 155--176.

\bibitem{lmms} F.~Lizzi, G.~Mangano, G.~Miele and G.~Sparano,
\emph{Constraints on unified gauge theories from noncommutative
geometry}, Mod.\ Phys.\ Lett.\ A 11 (1996) 2561--2572.

\bibitem{rw1} R.~Wulkenhaar, \emph{Graded differential Lie algebras
and model building}, to appear in J.\ Geom.\ Phys. (preprint
\texttt{hep-th/9607086})

\bibitem{cl} A.~Connes and J.~Lott, \emph{The metric aspect of
noncommutative geometry}, in: Proc.\ 1991 Carg\`ese Summer Conf.,
eds.\ J.~Fr\"ohlich et al, Plenum (New York 1992) 53-93.

\bibitem{rw3} R.~Wulkenhaar, \emph{The standard model within
non-associative geometry}, Phys.\ Lett.\ B \textbf{390} (1997)
119--127.

\bibitem{rw4} R.~Wulkenhaar, \emph{Graded differential Lie algebras
and $\mathrm{SU(5)} \times \mathrm{U(1)}$-grand unification}, to
appear in Int.\ J.\ Mod.\ Phys.\ A (preprint \texttt{hep-th/9607237}).

\bibitem{rw5} R.~Wulkenhaar, \emph{$\mathrm{SO(10)}$-unification in
noncommutative geometry revisited}, preprint \texttt{hep-th/9804046}.

\bibitem{tk} T.~Krajewski, \emph{Classification of finite spectral
triples}, preprint \texttt{hep-th/9701081}.

\end{thebibliography}
\end{document}